\documentclass[twoside]{article}
\usepackage{fleqn,espcrc2,epsfig}
\title{Heavy Baryons and electromagnetic decays}
\author{I. Scimemi
\address{Departament de F\'{\i}sica Te\`orica, IFIC, 
  Universitat de Val\`encia -- CSIC\\ E-46100 Burjassot (Val\`encia), Spain}
\thanks{Talk presented at $4^{th}$ International Conference Hyperons, Charm
                and Beauty Hadrons
             Conference, Valencia June 2000.
I thank M.C. Ba\~nuls and  A. Pich for collaboration.
This work has been supported in part by the European Union 
TMR Network  
``EURODAPHNE'' (Contract No. ERBFMX-CT98-0169).
Report:IFIC/00--64.}
}
\renewcommand{\a}{\alpha}

\newcommand{\g}{\gamma}

\newcommand{\G}{\Gamma}
\renewcommand{\S}{\Sigma}
\newcommand{\La}{\Lambda}
\newcommand{\eps}{\epsilon}

\newcommand{\bea}{\begin{eqnarray}}
\newcommand{\eea}{\end{eqnarray}}
\newcommand{\beq}{\begin{equation}}
\newcommand{\eeq}{\end{equation}}
\newcommand{\nn}{\nonumber}
\newcommand{\fr}{\frac}
\newcommand{\hl}{\hline}

\newcommand{\ra}{\rightarrow}

\begin{document}
\begin{abstract}
In this  talk   I review the theory of  electromagnetic decays of the ground state baryon multiplets with one
heavy quark  calculated using  Heavy Hadron Chiral Perturbation Theory~\cite{nosv}.
The M1 and E2 amplitudes for 
$S^{*}\ra S \g $, $S^{*} \ra T \g$ and $S \ra T \g$
are separately  analyzed.
All M1 transitions are calculated up to  ${\cal O}(1/\Lambda_\chi^2)$.
The E2 amplitudes contribute at the same order for $S^{*}\ra S \g $, while
for $S^{*} \ra T \g$ they first appear at  
${\cal O}(1/(m_Q \Lambda_\chi^2))$
and for $S \ra T \g$ are completely negligible.
Once the loop contributions is considered,
relations among different decay amplitudes are derived. 
In ref.~\cite{nosv} it is shown   that  the   coupling of the photon to 
 light mesons is responsible   
of  a sizable enhancement of these decay widths.
Furthermore,  one can obtain an absolute prediction for 
$\G(\Xi^{0'(*)}_c\ra \Xi^{0}_c \g) $ and
$\G(\Xi^{-'(*)}_b\ra \Xi^{-}_b \g) $.

\vspace{1pc}
\end{abstract}
\maketitle
\section{Introduction}

In Heavy Hadron Chiral Perturbation Theory (HHCPT) one constructs 
an effective Lagrangian whose basic fields are heavy hadrons and 
light mesons~\cite{wise}-\cite{cho}.
In ref.~\cite{chogeo},
 the formalism is extended to include also electromagnetism.
In this talk  I  describe 
how,  using this formalism,  one can  
calculate 
the electromagnetic decay width 
of some baryons containing a $c$ or a $b$ quark.
The details  of this computation
 are reported  in ref.\cite{nosv} and here I limit  myself  to trace
 its guidelines.     
In order to classify  these baryons one observe that 
the light degrees of freedom in the ground  state  of a 
baryon  with one heavy quark can be either in
a $s_l=0$ or in a $s_l=1$ configuration.
The first one corresponds  to $J^P=\frac{1}{2}^+$ baryons, which 
are annihilated by $T_i(v)$ fields which transform
as a $\bar{\bf3}$ under the chiral $SU(3)_{L+R}$ and as a doublet under 
the HQET $SU(2)_{v}$.
In the second case, $s_l=1$,
the  spin of the heavy quark and the light
 degrees of freedom combine together to form
$J^P=3/2^+$ and $J=1/2^+$ baryons which are degenerate
in  mass  in the $m_Q\rightarrow \infty$ limit.
 The spin-$\frac{3}{2}$ ones are annihilated
by the Rarita-Schwinger field $S_{\mu}^{* ij}(v)$ while the spin-$\frac{1}{2}$
baryons are destroyed by the Dirac field $S^{ ij}(v)$. 
They transform as a {\bf 6} under
$SU(3)_{L+R}$ and as a doublet under $SU(2)_{v}$ and are symmetric in the $i$, $j$ indices.
I consider the decays $S^{*}\ra S \g $ and $S^{(*)} \ra T \g$.
For most of these decays the  available phase space is   small,
 so that  the  emission of a pion is suppressed or even forbidden 
and the electromagnetic process becomes relevant.  
Moreover these kinds of decays are getting measured~\cite{expcleo}.
In the case of $S^* \ra S \g$ 
 all contributions  up  to order  ${\cal O}(1/\La_\chi^2)$ are calculated for 
 M1 and E2 transitions.
All divergences and scale 
dependence  
can be absorbed in the redefinition of one  ${\cal O}(1/\Lambda_\chi)$
coupling  for each type of process (M1, E2).
Eliminating the unknown constants  it is possible to find relations among the
 amplitudes which are valid up to the considered order.
 An analogous calculation can be performed  for  $S^* \ra T \g$.
In this case, the E2 contribution has to be computed 
 up to order ${\cal O}(1/m_Q \La_\chi^2)$, implying 
the intervention of two new constants.
Finally  for  $S\ra T \g$ the M1 amplitude is calculated up to
order  ${\cal O}(1/\La_\chi^2)$, while the E2 contribution
is found to be extremely suppressed.
In the case  $S^{(*)}\ra T \g$    it exists a process 
which do not receive any contribution from local terms in the Lagrangian and therefore
its width is described  by a finite chiral loop calculation: 
$\G(\Xi^{0'(*)}_c\ra \Xi^{0}_c \g) $ (and analogously 
$\G(\Xi^{-'(*)}_b\ra \Xi^{-}_b \g) $).
In  the following I comment  these   results  and I
 refer to ref.~\cite{nosv} for the  formalism and for 
a more complete comparison with other results existing in literature.
A similar formalism   can be applied  to the study of the magnetic moments 
  of the same baryons~\cite{nos}.
\section{Results for $S^*\ra S \g $ decays}
\label{sec:S}

\begin{figure}[ht]
\begin{center}
\epsfig{file=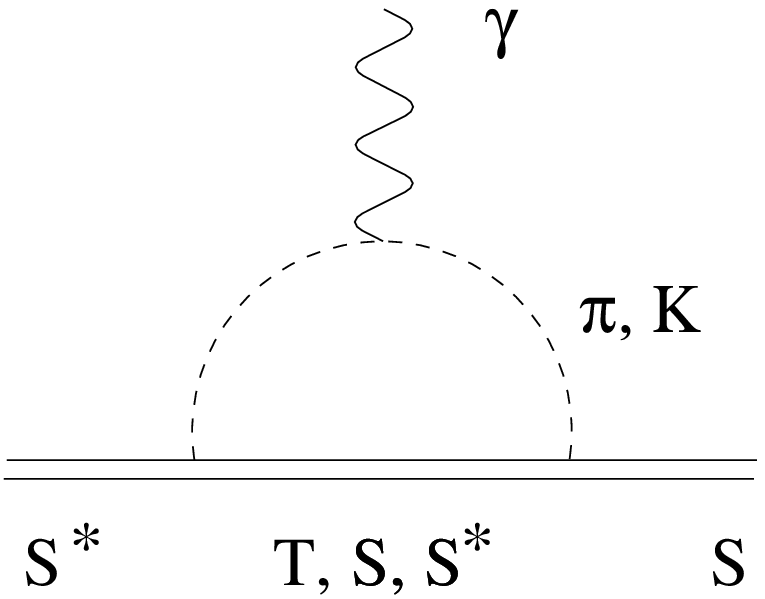,width=0.5\linewidth,angle=0}
\caption{Meson loops contributing to $S^*\ra S \g$.}
\label{fig:S}
\end{center}
\end{figure}

The decay amplitudes are decomposed  by
\beq
{\cal A}\left( B^*\to B \ \g\right) \, = \,
A_{M1}  \, {\cal{O}}_{M1} + A_{E2}  \, {\cal{O}}_{E2} \, ,
\eeq
where the corresponding M1 and E2 operators  are defined by
\bea
{\cal{O}}_{M1}&=& e \, \bar{B}\g_\mu \g_5   B^*_\nu\ F^{\mu\nu},  
\nn \\
{\cal{O}}_{E2}&=& i \, e \, \bar{B}\g_\mu \g_5   B^*_\nu\ v_\a \,
(\partial^\mu F^{\a\nu}+\partial^\nu F^{\a\mu}) \, ,
\label{eq:magop}
\eea
The leading  contributions to M1 transitions come from the
light-- and heavy--quark magnetic interactions  which are of 
 ${\cal O}(1/\La_\chi) $ and  ${\cal O}(1/m_Q) $, respectively.
We have  computed the next-to-leading chiral corrections of  
${\cal O}(1/\La_\chi^2) $,  which originate from the loop diagrams 
shown in fig.~\ref{fig:S}.

\begin{table}
\begin{center} 
\begin{tabular}{|c|c|c|}  \hl
c quark  &   $a_\chi$   & $a_{g_2}$  
\\ \hl
$\Sigma^{++*}_{c}\ra\Sigma^{++}_{c}\g $
& 2& $I_\pi$ + $I_K$  
\\
$\Sigma^{+*}_c\ra \Sigma^{+}_c \g$&
 1/2& $I_K/2$ 
\\
$\Sigma^{0*}_c \ra \Sigma^{0}_c\g $&
$-1$ &$-I_\pi$  
\\
$\Xi^{0'*}_c \ra \Xi^{0'}_c\g $    & $-1$&
 $-(I_\pi+I_K)/2$
\\
$\Xi^{+'*}_c \ra \Xi^{+'}_c \g $    & 1/2&
$I_\pi/2$ 
\\
$\Omega^{0*}_c\ra \Omega^{0}_c \g $ &$-1$
&  $-I_K$ 
\\ \hl
\end{tabular} 
\caption{Contributions to  M1 amplitudes for $S^* \ra S \g$. 
The values  of $a_{g_3}$  can be deduced from the ones of $a_{g_2}$ 
with the substitution
$I_i\ra m_i/m_K$ ($i=\pi,K$). }
\label{tab:uno}
\end{center} 
\end{table}
The resulting M1 amplitudes  can be written as:
\bea
A_{M1} (B^{*})& =& \fr{1}{\sqrt{3}} \left( - \fr{Q_{Q}}{m_Q} -
\fr{2 c_s}{3 \Lambda_\chi}\, a_{\chi}(B^{*}) \right. \nn \\
& &  + g_2^{2} \,\fr{\Delta_{ST}}{4 (4 \pi f_\pi)^2}\, a_{g_2}(B^{*})
 \nn \\
& & \left.+
g_3^{2} \,\fr{m_K}{4 \pi f_\pi^2}a_{g_3}\, (B^{*}) \right) \ .
\label{eq:magS}
\eea
In Table~\ref{tab:uno} we show the values of the coefficients $a_i(B^*)$ 
for the decays of baryons containing one charm or bottom quark.
In the  table,
\bea
\label{eq:intr}
I_i & \equiv & I(\Delta_{ST}, m_i) = 2 \left(-2 +\log{\fr{m_i^{2}}{\mu^{2}}}
\right) \nn \\ & & 
+ 2 \, {\sqrt{\Delta_{ST}^{2}- m_i^{2}}\over \Delta_{ST}}\,
 \nn \\ & & 
\times \log{\left(\fr{\Delta_{ST}+ \sqrt{\Delta_{ST}^{2}- m_i^{2}} }{ \Delta_{ST}- 
\sqrt{\Delta_{ST}^{2}- m_i^{2}}
 }\right)} \ . 
\eea
where $\Delta_{ST}$  is the mass difference between $S$ and $T$--baryons.
Due to flavor symmetry,  all contributions are equal 
for charm and bottom baryons, with the only exception of 
the term proportional to  the heavy  quark electric charge
($Q_{c}=+2/3$, $Q_{b}=-1/3$).
The main things to be observed are the following:
\begin{itemize}
\item
the  corrections  proportional to $g_2^{2}$ are obtained performing a 
one--loop integral (fig.~\ref{fig:S} with an $S$ baryon running in the loop) 
that has to be renormalized.
 It can be demonstrated~\cite{nosv} that
the scale $\mu$ dependence of the loop integrals is exactly canceled
by the corresponding dependence of the coefficient $c_S(\mu)$;
\item
the contribution proportional to $g_3^{2}$ involves  a loop integral
with a baryon of the $T$ multiplet running in the loop.
Since the Lagrangian does not have any mass term for $T$ baryons,
the result of the integral is convergent and proportional to  
the mass of the light mesons.
\end{itemize}
Looking at table~\ref{tab:uno} one sees that relations  among 
 the  decay amplitudes in which
 all unknown constants are eliminated can be easily found.
A complete list of them is reported in  ref.~\cite{nosv}.

The M1 and E2 amplitudes have identical $SU(3)$ structure.
The only difference is that   there are no $1/m_Q$ terms contributing to E2.
 Therefore,
one can construct for the  E2 amplitudes 
exactly the same relations as in the M1 case. 

The  E2 amplitudes  come at higher chiral order with respect to the M1 ones.
Therefore, the E2 contribution to the total width is suppressed  by  a 
factor $(E_\g/\La_\chi)^2\sim 5\%$.
In principle, it should be possible to determine experimentally  the
ratio $A_{E2}/A_{M1}$ by studying the angular distribution of photons
from the decay of polarized baryons \cite{savanu,RB:67,BSS:93}.
The Fermilab E-791 experiment has reported \cite{E791}
a significant polarization effect on the production of $\Lambda_c$ baryons,
which perhaps could be useful in future measurements of these
electromagnetic decays.
 In ref.~\cite{nosv} it has been observed also  that the  loop contribution 
can strongly enhance   the  decay widths.   In other words
 the coupling  of the photon to light  meson  can give the main
  contribution to the decay widths.

\section{Results for $S^* \ra T \g$ decays}
\label{sec:T}

The M1 and E2 operators for these decays are defined as in Eq.~(\ref{eq:magop}).
Similarly to what we have done in the previous paragraph, we write
the M1 amplitude for $S^* \ra T \g$ decays as
\bea
A_{M1} (B^*)& =& 
 -\sqrt{2}\, \fr{ c_{ST}}{ \Lambda_\chi } \, a_{\chi}(B^*) 
\nn \\ & & +
 g_2\, g_3 \,\fr{\Delta_{ST}}{2 \sqrt{2} (4 \pi f_\pi)^2}\, a_{g}(B^*) \ .
\label{eq:amt}
\eea
The  value of  the parameters entering this equation can be found  in ref.~\cite{nosv}.  The final result do not depend on  the  heavy quark mass or charge.
All constants  can be eliminated in the relations 
\bea
A_{M1}(\S^{+*}_c)-A_{M1}(\Xi^{+'*}_c)& =&  -3\, A_{M1}(\Xi^{0'*}_c)\ , \nn \\
A_{M1}(\S^{0*}_b)-A_{M1}(\Xi^{0'*}_b)& =&  -3\, A_{M1}(\Xi^{-'*}_b) \ .
\label{eq:innom}
\eea

It is interesting to notice that $A_{M1}(\Xi^{0'*}_c)$ does not depend 
on $c_{ST}$.
Since at ${\cal O}(1/\Lambda_{\chi}^2)$
this decay does not get any contribution from local terms,
its M1 amplitude results from a {\it finite} chiral loop calculation
(it cannot be divergent because there is no possible counter-term to
renormalize it), so that we have an absolute prediction for its value
in terms of $g_2$ and $g_3$.
Using  the experimental value of  $g_3$~\cite{gural,cleo}  and the
 corresponding  value of $g_2$~\cite{gnec} 
  derivable from the quark model,
one finds (see also ref.~\cite{wald})
\bea
\G_{M1}(\Xi^{0'*}_c)& = & 5.1 \pm 2.7 \ {\rm keV} \nn \\
\G_{M1}(\Xi^{-'*}_b)& = & 4.2 \pm 2.4 \ {\rm keV}
\label{eq:xic}
\eea
where  the dominant error come from the uncertainty on $g_{2,3}$.

The E2 amplitude in  $S^*\to T\gamma$ 
is suppressed by an extra power of $1/m_Q$.
The first non-zero contributions comes at  
${\cal O}(1/m_Q \Lambda_{\chi}^2)$.
It is important to note that at this order it appears an operator,
 which break spin symmetry,
\bea
{\cal{L}}^{\prime} &=& i\, \fr{g'}{m_Q}\,
\left [\epsilon_{ijk} \,\bar{T}^i\sigma^{\mu\nu} (\xi_{\mu})_l^j S_{\nu}^{kl}
\right. \nn \\ 
& & \left. +
\epsilon^{ijk} \,\bar{S}_{kl}^{\mu} \sigma_{\mu\nu} (\xi^{\nu})_j^l T_i\right ] \ ,
\label{eq:nova}
\eea
which gives rise to divergent loop diagrams.  Moreover finite contributions
 of the same order   come from 
\beq
-i \fr{c^{E2}_T}{m_Q \Lambda_{\chi}^2} \,\eps_{ijk}\,
\bar{T}^i\sigma_{\mu\nu} Q_l^j S_{\a}^{kl} 
\,\partial^{\a} \tilde{F}^{\mu \nu} \ .
\label{eq:counte}
\eeq
Both  the  contributions coming from eq.~\ref{eq:nova}--\ref{eq:counte}
 where not considered before in literature.

By eliminating the unknown coupling constants,
one  can deduce the relation
\beq
A_{E2}(\S^{+*}_c)-A_{E2}(\Xi^{+'*}_c) =  -3 \, A_{E2}(\Xi^{0'*}_c)
\ .
\eeq
The same relation holds for the corresponding $b$ baryons, since
\beq
A_{E2}(B^*_b)=\fr{m_c}{m_b} \, A_{E2}(B^*_c) \ .
\eeq

The decays $\Xi^{0*}_c \ra \Xi^{0}_c \g$   and $\Xi^{-*}_b \ra \Xi^{-}_b \g $
do not get any contribution from the local term proportional to
$c^{E2}_T$; their $O(1/m_Q\Lambda_\chi^2)$ E2 amplitude is
also given by a finite loop calculation. Unfortunately,
since the coupling $g'$ is not known, there is no absolute prediction
in this case.
An experimental measurement of these E2 amplitudes 
would provide a direct estimate of $g'$.

\section{Results for $S \ra T \g$}
\label{sec:st}

The calculation of the M1 amplitude for $S \ra T \g$ decays is analogous 
to that of the previous section.
Now the M1 operator is defined as
\beq
{\cal O}_{M1}=i e \,\bar{B}_T \sigma_{\mu \nu} B_S \, F^{\mu \nu} 
\eeq
and the corresponding amplitude can be written in the form
\beq
A_{M1} (B) = 
\fr{1}{\sqrt{6}} \fr{ c_{ST}}{ \Lambda_\chi }\, a_{\chi}(B)-
g_2 g_3 \,\fr{\Delta_{ST}\, a_{g}(B)}{4 \sqrt{6} (4 \pi f_\pi)^2 } \ ,
\label{eq:ams}
\eeq
where the coefficients satisfy
\beq
a_{\chi}(B)=a_{\chi}(B^*) , \qquad\quad a_{g}(B)=a_{g}(B^*) \ .
\eeq
Therefore, the relation~(\ref{eq:innom}) is also valid in this case.
The widths of the decays
$\Xi^{0'}_c \ra \Xi^{0}_c \g$  and $\Xi^{-'}_b \ra \Xi^{-}_b \g $
can  be predicted through a finite loop calculation. 
From
\beq
\G (S\ra T \g )=16 \a_{em}\,\fr{ E_\g^3 M_T}{  M_{S}} \, |A_{M1}|^2 \ ,
\eeq
we find
\bea
\Gamma(\Xi^{0'}_c)&=& (1.2\pm 0.7) \; {\rm KeV} \ , \nn \\
\Gamma(\Xi^{-'}_b)&=& (3.1\pm 1.8) \; {\rm KeV} \ .
\label{eq:xi}
\eea
%

Again the dominant  error in Eq.~(\ref{eq:xi}) is given by the uncertainty of 
$g_{2,3}$.

For these decays the E2 amplitude is further suppressed than in the previous cases. 
The lowest--order contribution appears at 
${\cal O} (1/m_Q^3 \Lambda_{\chi}^2)$ and, therefore, can be neglected.

\end{document}